\RequirePackage{lineno}
\documentclass[twocolumn,aps,prc,showpacs,superscriptaddress,preprintnumbers,floatfix,nofootinbib]{revtex4}
\usepackage{epsfig,graphics}
\usepackage{graphicx}
\usepackage{dcolumn}
\usepackage{bm}
\usepackage{amsmath}
\usepackage[usenames]{color}
\usepackage{ulem} 

\begin{document}

\title{Medium modifications of photon-tagged jet fragmentation function \\in high-energy heavy-ion collisions}

\author{Guo-Liang Ma}
\affiliation{Shanghai Institute of Applied Physics, Chinese Academy of Sciences, Shanghai 201800, China}


\begin{abstract}
Based on a multiphase transport model, medium modifications of prompt photon-tagged jet fragmentation function are investigated by comparing prompt photon-tagged hadron azimuthal correlation in Au+Au collisions (0-40\%) with that in p+p collisions at $\sqrt{s_{_{\rm NN}}}$ = 200 GeV. The measured modification factor, $I_{AA}$, increases with the increasing integration range of the away side, which reveals a medium-modified jet shape in which the medium enhancement of soft particles is preferentially located far away from the jet axis. The $I_{AA}$ largely results from strong interactions between jets and partonic matter. However, both hadronization of coalescence and hadronic rescatterings play certain roles to modify the $I_{AA}$. These behaviors reflect a dynamical evolution of modifications of the prompt photon-tagged jet fragmentation function in high-energy heavy-ion collisions.

\end{abstract}


\maketitle

\section{Introduction}
\label{sec:intro}
Plentiful experimental results disclose that a deconfined quark-gluon plasma (QGP) could be created by high-energy heavy-ion collisions at the Relativistic Heavy Ion Collider (RHIC) and the Large Hadron Collider (LHC)~\cite{Adams:2005dq, Adcox:2004mh}. Jets, produced back-to-back in the initial hard scatterings, serve as important hard probes to study the new formed matter, because they lose energy when passing through the matter~\cite{Wang:1991xy}. Much understanding of the properties of new partonic matter has been gained from jet-related measurements, especially from dijet observations such as the disappearance of back-to-back jets~\cite{Adler:2002tq} and dijet transverse momentum ($p_{T}$) asymmetry~\cite{Aad:2010bu,Chatrchyan:2011sx}. However, a dijet is a limited probe, because it can not probe the hot medium very deeply due to the surface bias~\cite{Zhang:2007ja}. On the other hand, a prompt photon-jet, i.e. a $\gamma$-jet, is thought to be a golden probe with many advantages, since photons escape the medium without interacting with the medium via the strong force~\cite{Zhang:2009rn}. For instance, high-$p_{T}$ prompt photons can provide a natural calibration of initial jet energy, and thus the recent LHC experimental results about photon-jet $p_{T}$ imbalance present direct and less biased quantitative measures of jet energy loss in the hot medium~\cite{Chatrchyan:2012gt,ATLAS:2012cna}, which suggest that jets significantly lose energy due to strong final-state interactions~\cite{Dai:2012am, Wang:2013cia, Qin:2012gp}.  Compared to dihadron azimuthal correlation, photon-tagged hadron azimuthal correlation has another advantage of a constant combinatorial background only, because high-$p_{T}$ photons do not flow~\cite{Adare:2011zr}. Owing to these advantages of photons, it becomes possible to approximately measure the jet fragmentation function and jet shape through photon-tagged hadron azimuthal correlation, which also saves people trouble of dealing with the complex jet reconstruction in the transitional measurements~\cite{Chatrchyan:2012gw,CMS:2012wxa,ATLAS:2012ina}. Recently, PHENIX measured the jet fragmentation function by means of direct photon-hadron azimuthal correlations; the measured jet fragmentation function ratio of Au+Au to p+p, $I_{AA}$, appears suppressed at low $\xi=ln(1/z)$ and enhanced at high $\xi$~\cite{Adare:2012qi}, which is consistent with the theoretical calculations from the Borghini and Weidemann's modified leading log approximation (BW-MLLA)~\cite{Borghini:2005em} and and the yet another jet energy loss model (YaJEM)~\cite{Renk:2011qf} based on jet radiation energy loss mechanisms. It indicates that the lost energy has been redistributed into enhanced production of soft particles to modify the jet shape. In this work, medium modifications of the prompt photon-tagged jet fragmentation function are investigated via prompt photon-tagged hadron azimuthal correlations in p+p and Au+Au collisions at $\sqrt{s_{_{\rm NN}}}$ = 200 GeV within a multiphase transport (AMPT) model, which includes both dynamical evolutions of partonic and hadronic phases. The measured modification factors, $I_{AA}$, for different integration ranges of the away side are qualitatively reproduced by the AMPT model, which indicates a medium-modified jet shape in which the medium enhancement of soft particles is preferentially located far away from the jet axis. The dynamical evolution of the $I_{AA}$ for different stages in high-energy heavy-ion collisions is presented. 

\section{The AMPT Model}
\label{sec:model}

The AMPT model with string-melting scenario is employed in this work~\cite{Lin:2004en}. It consists of four main stages of high-energy heavy-ion collisions: the initial condition, parton cascade, hadronization, and hadronic rescatterings. In order to study the energy loss behavior of a photon jet, a photon jet of $p_{T}^{\gamma}\sim$ 4 GeV/$c$ is triggered with the jet-triggering technique in HIJING~\cite{Wang:1991hta,Gyulassy:1994ew},  since the production cross section of photon jet is quite small especially for large $p_{T}^{\gamma}$. Three hard photon-jet production processes with high virtualities are taken into account in the initial condition, including $q+\bar{q}\rightarrow g+\gamma$, $q+\bar{q} \rightarrow \gamma+\gamma$, and $q+g \rightarrow  q+\gamma$~\cite{Sjostrand:1993yb}. The symbol $\gamma$ hereinafter denotes the prompt photons from the hard processes. For these $\gamma$, their birth information is kept, since they only participate in weak electromagnetic interactions. However, the high-$p_{T}$ primary partons evolve into jet showers full of lower virtuality partons through initial- and final- state QCD radiations. The jet parton showers are converted into clusters of on-shell quarks and anti-quarks through the string-meting mechanism of AMPT model. After the melting process, both a quark and anti-quark plasma and jet quark showers are built up. In the following, Zhang's parton cascade (ZPC) model~\cite{Zhang:1997ej} automatically simulates all possible elastic partonic interactions among medium partons and jet shower partons, but without including inelastic parton interactions or further radiations at present. When the partons freeze-out, they are recombined into medium hadrons or jet shower hadrons via a simple coalescence model. The final-state hadronic rescatterings between jet shower hadrons and hadronic medium can be described by a relativistic transport (ART) model~\cite{Li:1995pra}. The AMPT model has successfully given some good descriptions to some experimental observables, such as $p_{T}$ spectra~\cite{Lin:2004en}, harmonic flows~\cite{Chen:2004dv, Zhang:2005ni, Chen:2006ub, Han:2011iy}, $\gamma$-hadron~\cite{Li:2010ts}, and dihadron correlations~\cite{Ma:2010dv} at the RHIC energies. Consistently with the previous AMPT studies, a large partonic interaction cross section, 10 mb, is used to simulate the 0-40\% centrality bin in Au+Au collisions at $\sqrt{s_{_{\rm NN}}}$ = 200 GeV, while a zero partonic interaction cross section is used to turn off the partonic interactions in p+p collisions at $\sqrt{s_{_{\rm NN}}}$ = 200 GeV. Constant backgrounds are removed from prompt photon-hadron azimuthal correlations in p+p and Au+Au collisions by using a zero-yield-at-minimum (ZYAM) method~\cite{Ajitanand:2005jj}, since high-$p_{T}$ prompt photons do not flow in the AMPT simulations.

\section{Results and Discussions}
\label{sec:results}

\begin{figure}
\includegraphics[scale=0.5]{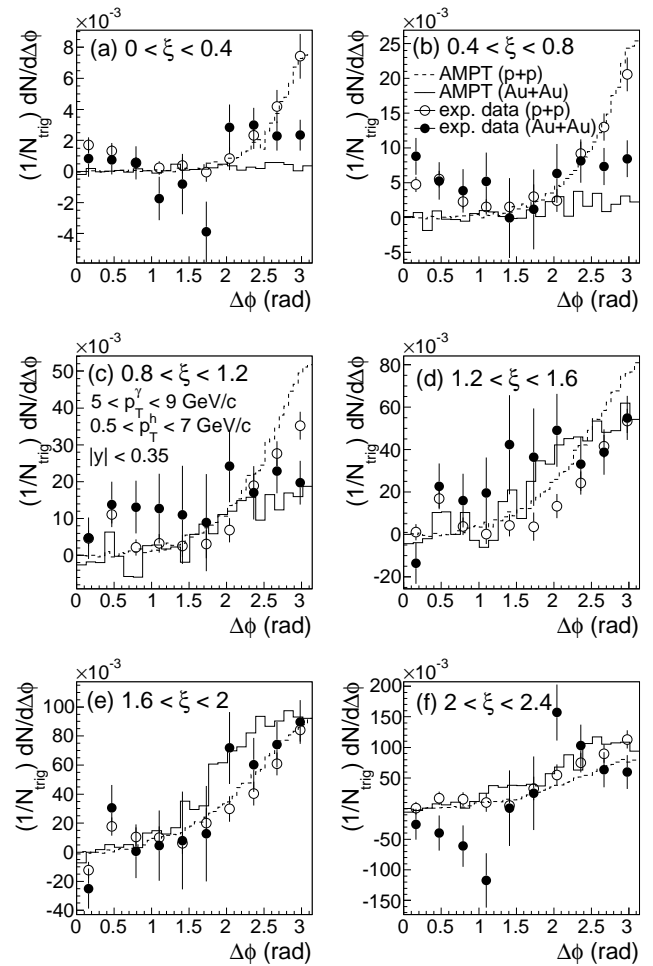}
\caption{(a)-(f) The $\gamma$-tagged hadron azimuthal correlations for different $\xi$ bins in p+p collisions and Au+Au collisions (0-40\%) at $\sqrt{s_{_{\rm NN}}}$ = 200 GeV. The AMPT results are shown by histograms, while experimental data are shown by circles for which error bars indicate the statistical uncertainties only and the systematic uncertainties are not included~\cite{Adare:2012qi}.}
 \label{fig-phidis}
\end{figure}

Figure~\ref{fig-phidis} (a)-(f) present the AMPT results on constant-background-removed azimuthal angle ($\Delta\phi=\phi_{h}-\phi_{\gamma}$) correlations between prompt photons ($5 < p_{T}^{\gamma} <$ 9 GeV/$c$) and final hadrons ($0.5 < p_{T}^{h} <$ 7 GeV/$c$) within a mid-rapidity range ($|y| <$ 0.35) for different $\xi$ bins in p+p collisions and Au+Au collisions (0-40\%) at $\sqrt{s_{_{\rm NN}}}$ = 200 GeV, in comparison with experimental data~\cite{Adare:2012qi}. The variable $\xi = ln(1/z)$, where $z = p_{T}^{h}/p_{T}^{\gamma}$ which is the ratio of the associated hadron transverse momentum, $p_{T}^{h}$, to the photon transverse momentum, $p_{T}^{\gamma}$. For p+p collisions, the AMPT model roughly can describe the data for all different $\xi$ bins. However, for the $\gamma$-tagged hadron azimuthal correlations in Au+Au collisions, the AMPT model gives good descriptions for high-$\xi$ bins, but underestimates the data for low-$\xi$ bins.

\begin{figure}
\includegraphics[scale=0.45]{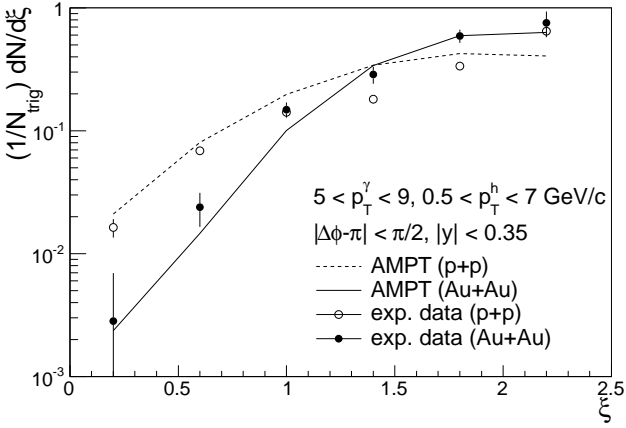}
\caption{The $\gamma$-tagged jet fragmentation functions for p+p collisions and Au+Au collisions (0-40\%) at $\sqrt{s_{_{\rm NN}}}$ = 200 GeV. The AMPT results are shown by curves, while experimental data are shown by circles for which error bars indicate the statistical uncertainties only and the systematic uncertainties are not included~\cite{Adare:2012qi}. }
 \label{fig-fragfunc}
\end{figure}

Since a high-$p_{T}$ prompt photon can give a good approximation to the initial momentum for the away-side jet, a $\gamma$-tagged jet fragmentation function, ($1/N_{trig})dN/d\xi$, can be obtained by integrating the away side of $\gamma$-tagged hadron azimuthal correlation [($1/N_{trig})dN/d\Delta\phi$] for different $\xi$ bins within a $\Delta\phi$ range. Figure~\ref{fig-fragfunc} shows the $\gamma$-tagged jet fragmentation functions for full away side ($|\Delta\phi-\pi| < \pi/2$) in p+p collisions and Au+Au collisions (0-40\%) at $\sqrt{s_{_{\rm NN}}}$ = 200 GeV. The AMPT model roughly can describe the experimental data for both p+p and Au+Au collisions.

\begin{figure}
\includegraphics[scale=0.45]{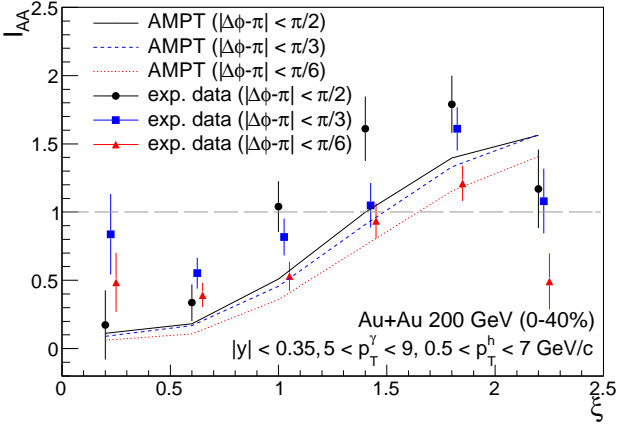}
\caption{(Color online) The $I_{AA}$ for full away side ($|\Delta\phi-\pi| < \pi/2$) and for two restricted away-side integration ranges ($|\Delta\phi-\pi| < \pi/3$ and $|\Delta\phi-\pi| < \pi/6$). The AMPT results are shown by curves, while experimental data are shown by symbols for which error bars indicate the statistical uncertainties only and the systematic uncertainties are not included~\cite{Adare:2012qi}. Some points are slightly shifted for clarity.
}
 \label{fig-iaa}
\end{figure}

To study medium modifications of the jet fragmentation, Figure~\ref{fig-iaa} presents the AMPT results on the ratios $I_{AA}$ of jet fragmentation functions in Au+Au collisions to those in p+p collisions for different away-side integration ranges, in comparison with experimental data. The AMPT results can qualitatively describe the data, which show a strong suppression at low $\xi$ and a large enhancement at high $\xi$. It implies a decrease of hard associated particle yield and an increase of soft associated particle yield for away-side in Au+Au collisions, relative to in p+p collisions.  With increasing integration range of the away side, the $I_{AA}$ from AMPT simulations increases, especially in the high-$\xi$ range, which is similar to the trend found in the experimental results. However, the AMPT results show a much smaller angular dependence than the data. The angular dependence displays the modifications of $\gamma$-tagged jet shape in the $\Delta\phi$ direction, which reflects that the medium enhancement of soft particles is preferentially located at large azimuthal angle which is far away from the jet axis. These behaviors are also consistent with the modifications of reconstructed-jet shape in Pb+Pb collisions at the LHC energy~\cite{CMS:2012wxa,Ma:2013uqa,Chatrchyan:2013kwa}.

\begin{figure}
\includegraphics[scale=0.5]{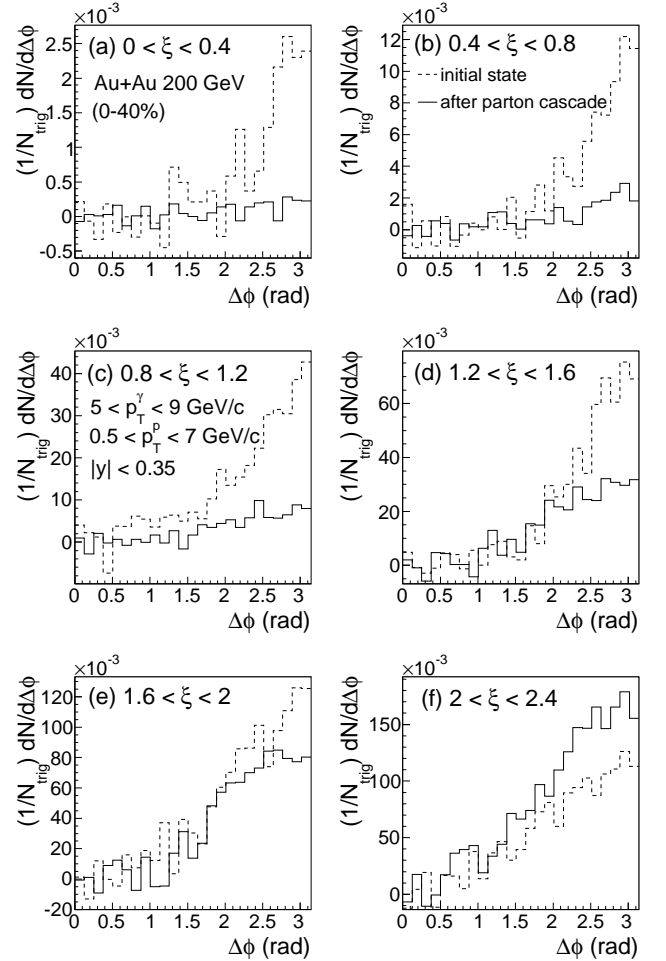}
\caption{(a)-(f) The $\gamma$-tagged parton azimuthal correlations for in the initial state (dash) and after the process of parton cascade (solid) for different $\xi$ bins in Au+Au collisions (0-40\%) at $\sqrt{s_{_{\rm NN}}}$ = 200 GeV. 
}
\label{fig-zpceff}
\end{figure}

\begin{figure}
\includegraphics[scale=0.45]{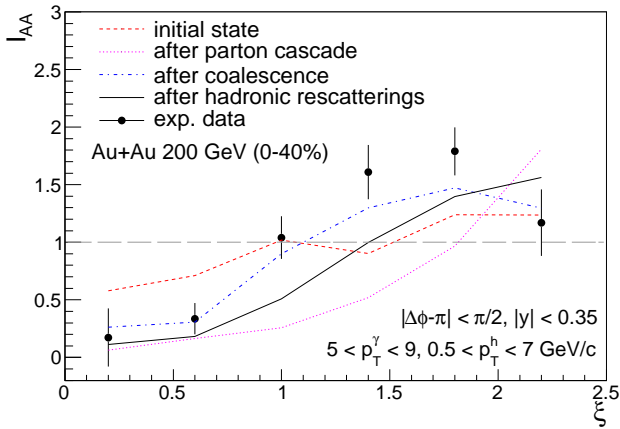}
\caption{(Color online) The $I_{AA}$ for full away side ($|\Delta\phi-\pi| < \pi/2$) at or after different evolution stages. The AMPT results are shown by curves, while experimental data are shown by circles for which error bars indicate the statistical uncertainties only and the systematic uncertainties are not included~\cite{Adare:2012qi}.
}
\label{fig-iaaevo}
\end{figure}

Because heavy-ion collisions actually are dynamical evolutions including many important stages, the stage evolution of $\gamma$-tagged correlation can provide us more important information about the picture of jet energy loss. Figures~\ref{fig-zpceff} (a)-(f) show the AMPT results on the $\gamma$-tagged parton azimuthal correlations between a prompt $\gamma$ ($5 < p_{T}^{\gamma} <$ 9 GeV/$c$) and partons ($5 < p_{T}^{p} <$ 9 GeV/$c$) within a mid-rapidity range ($|y| <$ 0.35) for different $\xi$ bins in the initial state and after the process of parton cascade in Au+Au collisions (0-40\%) at $\sqrt{s_{_{\rm NN}}}$ = 200 GeV. It can be seen that the $\gamma$-tagged parton correlation is suppressed for $0 < \xi < 2$ [Figure~\ref{fig-zpceff} (a)-(e)], but enhanced for $2 < \xi < 2.4$ [Figure~\ref{fig-zpceff} (f)] after the process of parton cascade. Quantitatively, Figure~\ref{fig-iaaevo} gives the AMPT results on the $I_{AA}$ as functions of $\xi$ at or after different evolution stages. The initial $I_{AA}$ is close to unity, which shows less modifications mainly due to the nuclear effects in the initial state of heavy-ion collisions~\cite{Wang:1991hta,Gyulassy:1994ew}. However, the $I_{AA}$ becomes strongly suppressed at low $\xi$ and largely enhanced at high $\xi$ after the process of parton cascade, which indicates that away-side jets lose energy significantly and excite the medium partons via the strong interactions between jets and the partonic medium.  The hadronization of coalescence weakens the formed modifications by increasing the $I_{AA}$ at low and intermediate $\xi$ but decreasing it at very high $\xi$, as the coalescence always combines high-$\xi$ partons into low-$\xi$ hadrons. It has been argued that there may exist a competition between coalescence and fragmentation for jet hadronization in high-energy heavy-ion collisions in my previous work~\cite{Ma:2013gga}, but only the coalescence mechanism is considered here for simplicity. Finally, the process of hadronic rescatterings slightly strengthens the modifications by decreasing the $I_{AA}$ at low and intermediate $\xi$ and increasing it at very high $\xi$, owing to the softening effects of continuous hadronic scatterings between jets and hadronic medium and resonance decays. The process of hadronic rescatterings makes the description of $I_{AA}$ worse compared to that after coalescence. It is checked that a better angular dependence could be obtained if switching off hadronic rescatterings. However, hadronic rescatterings are a very essential stage in the AMPT model especially for increasing the total multiplicity of final hadrons (d$N_{ch}$/d$\eta$) to match the experimental data. It should be mentioned that a similar effect of hadronic rescatterings on a reconstructed-jet shape is also observed in the AMPT simulations of Pb+Pb collisions at the LHC energy~\cite{Ma:2013uqa}. Therefore, the jet shape modification in hadronic medium certainly deserves further study for a satisfactory understanding of experimental data. In general, these results show the dynamical evolution behaviors of the $\gamma$-tagged jet fragmentation function in Au+Au collisions at the RHIC energy.

\hspace{1mm}
\section{Summary}
\label{sec:summary}

In summary, the medium modifications of $\gamma$-tagged jet fragmentation functions are investigated via $\gamma$-tagged hadron azimuthal correlations in p+p and Au+Au collisions at $\sqrt{s_{_{\rm NN}}}$ = 200 GeV within a multiphase transport model. Owing to strong interactions between away-side jets and the partonic medium, jets lose energy significantly and excite the medium which brings basic modification features to the $\gamma$-tagged jet fragmentation function ratio of Au+Au to p+p, $I_{AA}$. The hadronization of coalescence weakens the modifications, while hadronic rescatterings strengthens the modifications. It should be mentioned that my simulations currently do not include the mechanism of jet radiation energy loss, but a large partonic interaction cross section can partially play an effective role. In addition, my previous study found that the $\gamma$-jet-tagged hadron azimuthal correlation with a restricted $\gamma$-jet $p_{T}$ imbalance ratio is a detail tomography tool to see the medium responses to different initial $\gamma$-jet-production configurations~\cite{Ma:2013bia}. In the future, it would be interesting to see the medium modifications of $\gamma$-tagged jet fragmentation functions for different initial $\gamma$-jet-production configurations with the help of $\gamma$-jet $p_{T}$ imbalance at the RHIC and LHC energies.

\section*{ACKNOWLEDGEMENTS}

The author is grateful to Professor Barbara Jacak for calling his attention to this topic, and the HIRG PC farm for computer time. This work was supported by the Major State Basic Research Development Program in China under Contract No. 2014CB845404, the NSFC of China under Projects No. 11175232, No. 11035009, and No. 11375251, the Knowledge Innovation Program of CAS under Grant No. KJCX2-EW-N01, the Youth Innovation Promotion Association of CAS, the project sponsored by SRF for ROCS, SEM, CCNU-QLPL Innovation Fund under Grant No. QLPL2011P01, and the ``Shanghai Pujiang Program" under Grant No. 13PJ1410600.

\end{document}